# From Dialect Gaps to Identity Maps: Tackling Variability in Speaker Verification

*Abdulhady Abas Abdullah[1]*, Abdulhady.abas@ukh.edu.krd;*

Soran Badawi[2] soran.sedeeq@chu.edu.iq;

Dana A. Abdullah[3] Dana.abdullah531@gmail.com;

Dana Rasul Hamad[4] dana.hamad@soran.edu.iq;

[1] Artificial Intelligence and Innovation Centre, University of Kurdistan Hewlér, Erbil, Kurdistan Region, Iraq

[2]Language Center, Charmo University, Chamchamal, Kurdistan Region, Iraq

[3]Faculty of Engineering and Computer Science, Qaiwan International University, Sulaymaniyah, KRG, Iraq

[4]Computer Science Department, Faculty of Science, Soran University, Soran, Erbil, Kurdistan Region, Iraq

* Corresponding author: Abdulhady.abas@ukh.edu.krd

**Abstract**

The complexity and difficulties of Kurdish speaker detection among its several dialects are investigated in this work. Because of its great phonetic and lexical differences, Kurdish with several dialects including Kurmanji, Sorani, and Hawrami offers special challenges for speaker recognition systems. The main difficulties in building a strong speaker identification system capable of precisely identifying speakers across several dialects are investigated in this work. To raise the accuracy and dependability of these systems, it also suggests solutions like sophisticated machine learning approaches, data augmentation tactics, and the building of thorough dialect-specific corpus. The results show that customized strategies for every dialect together with cross-dialect training greatly enhance recognition performance.

## 1. Introduction

Speaker recognition is a biometric method based on analysis of particular traits obtained from a person's voice, therefore authenticating their identification. By analyzing the distinct characteristics of their voice signals, this automated approach detects the speaker. These vocal utterances are used by the speaker recognition system to verify speaker identification and govern access to services such voice calling, voicemail, and security control (Kabir et al., 2021). These technologies enable machines to precisely interpret and alter spoken language, thereby enabling voice assistants, transcription solutions, and openness tools to appropriately reflect varied applications (Kheddar et al., 2023).

As Figure 1 shows, the speech recognition process may be divided up into a number of important phases. It starts with audio input, usually spoken words or phrases caught by a recording device. Following that, procedures like filtering and noise reduction help to pre-process the original audio stream thereby enhancing its quality. After that, feature extraction helps the audio data to be in a more controllable form by identifying and extracting important properties. Mel Frequency Cepstral Coefficients (MFCC), computed with filter banks intended to replicate human voice perception, are a commonly used method in this stage. These coefficients sample power spectra produced from the Short-Time Fourier Transform applied across overlapping frames produced by segmenting the audio input. Usually using Hidden Markov Models (HMM) or deep neural networks (DNN), the Acoustic Model is then used to build the link between the retrieved data and language units like phonemes or sub-word components (Yakubovskyi and Morozov, 2023). Concurrent with this end-to--end (E2E) approach, which avoids intermediate language units, directly connects acoustic features with words or phrases using advanced neural network architectures such as Recurrent Neural Networks (RNN), Convolutional Neural Networks (CNN), or Transformers (Vaswani, 2017). Furthermore included in the process are predictions from several models combined with transfer learning, a method wherein pre-training models on large-scale datasets are then fine-tuned for particular automated speech recognition (ASR) tasks with limited resources. Following this, a language model (Ferrer-Estévez and Chalmeta) calculates the likelihood of word sequences in a given language, therefore enhancing accuracy and narrowing recognition possibilities. At last, segmentation—a crucial process separating the continuous audio stream into meaningful units like phonemes, syllables, or words—prepares it for subsequent analysis and identification.

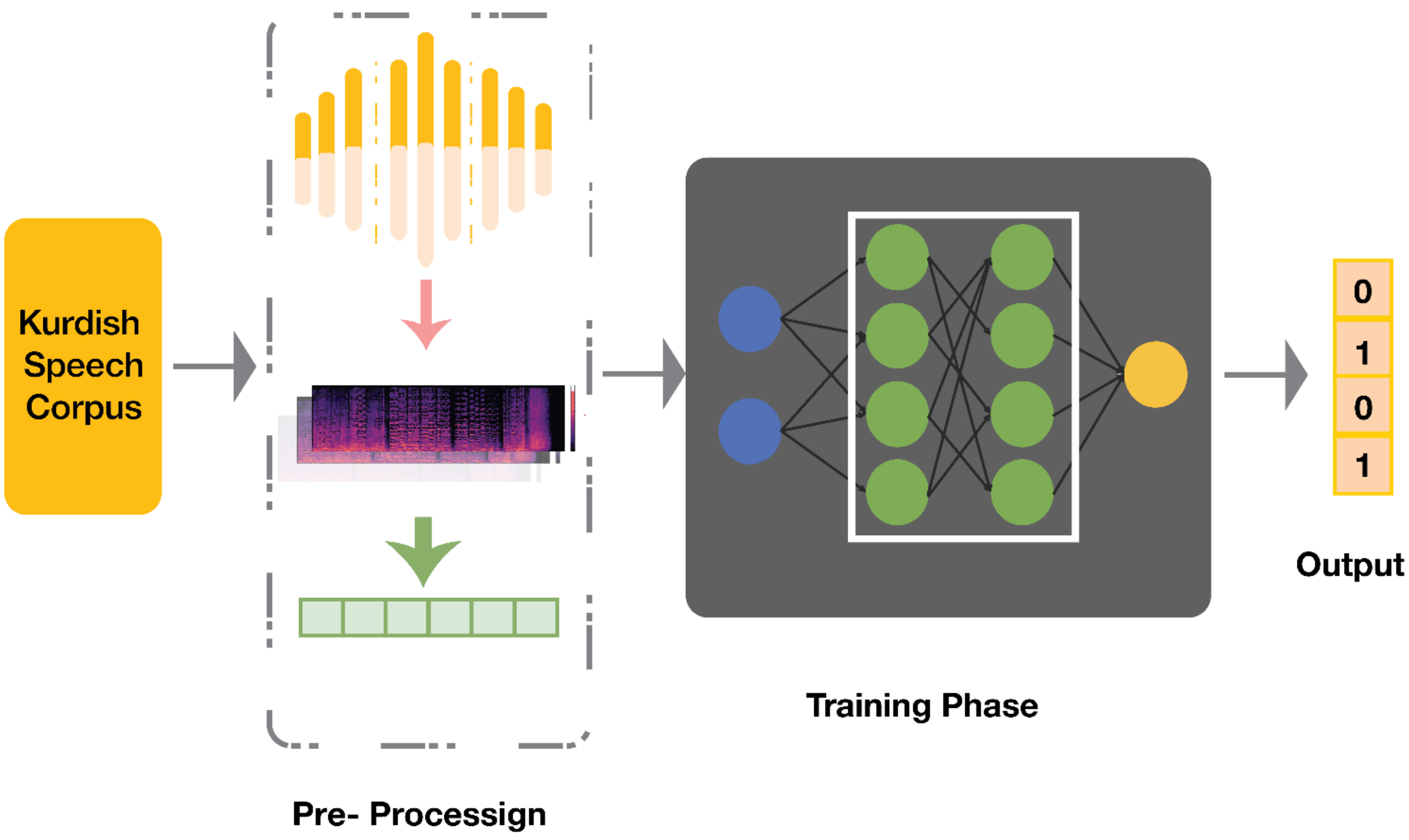

*Figure 1.Basic structure of an automatic speaker recognition system*

There are more than 40 million speakers of the Indo-European language known as Kurdish. (Badawi, 2023). It is composed of 33 letters and belongs to the Indo-Iranian branch of the Indo-European family. Kurdish exhibits parallels to Persian and is primarily spoken in Iran, Turkey, Iraq, and Syria. Notably, it is acknowledged as one of the official languages in Iraq. The language predominantly comprises two principal dialects: Central Kurdish (Sorani) and Northern Kurdish (Kurmanji). Additionally there are minor dialects like Zazaki, mostly spoken in Turkey, and Gorani (Hawrami), spoken by tiny groups in Iraq and Iran (Badawi et al., 2024). For speaker recognition study, the language offers a fascinating and challenging topic. It stands for notable dialectal variety. The dialectal variety among Kurds presents significant difficulties for systems of speaker identification. Variations in syntax, vocabulary, and pronunciation across the several dialects impede the growth of strong models equipped to handle cross-dialectal situations. Moreover, the scarcity of language resources and annotated datasets for Kurdish accentuates these difficulties, therefore speaker recognition in Kurdish dialects is an understudied field in both scholarly work and pragmatic uses (Abdullah et al., 2024). The possibility to introduce speaker identification systems in Kurdish-speaking areas is still mostly unrealized without focused research and solutions. Research on Kurdish dialectal variants is lacking in major proportion. This is a result of the limited computational language resources available for Kurdish dialects, which impede systematic study and technical advancement even further (Muhamad et al., 2024).

Inspecting and addressing the difficulties with speaker recognition across the several dialects of the Kurdish language is the major aim of this work. Mostly in cross-dialectal situations, this entails determining the linguistic and dialectal elements affecting the accuracy and dependability of speaker identification systems. Furthermore, the research intends to provide creative approaches and evaluate their success in tackling these difficulties. By doing this, this study aims to improve speaker identification technologies for Kurdish speakers, thereby promoting a more inclusive approach in computational linguistics and increasing the relevance of these systems in security, communication, and tailored services.

## 2. Literature Review

Speaker identification systems fall into several forms depending on their recognition criteria. The many approaches of speaker recognition are demonstrated in Figure 2. The acknowledged methods will be fully described in the ensuing subsections.

First level addresses speaker diarization, speaker verification, and automatic speaker identification. They are considered fundamental and effective methods for preventing unauthorized access to computer systems. Speaker identification establishes the identity of an anonymous speaker based on their spoken utterances by comparing their voice to a repository of pre-recognized voices. This approach functions as a 1: N match, where a particular utterance is compared to a variety of templates. In contrast, speaker verification aims to authenticate the identity of a specific individual who claims to be someone else. It correlates the characteristics of the voice that has been acquired with the features stored in a comprehensive voice model database for all speakers; however, speaker verification ties the acquired voice characteristics only to the speaker's claimed identity, resulting in a 1:1 match in which one person's speech is matched to a single template. Speaker diarization entails segmenting speech from multiple speakers into distinct sections associated with each individual and is a crucial component of speaker recognition systems. It finds applications in various areas, including video captioning and analysis of conversational content. The recognition criteria for these systems can either be text-dependent or text-independent, a topic that will be explored in the following section (Kabir et al., 2021).

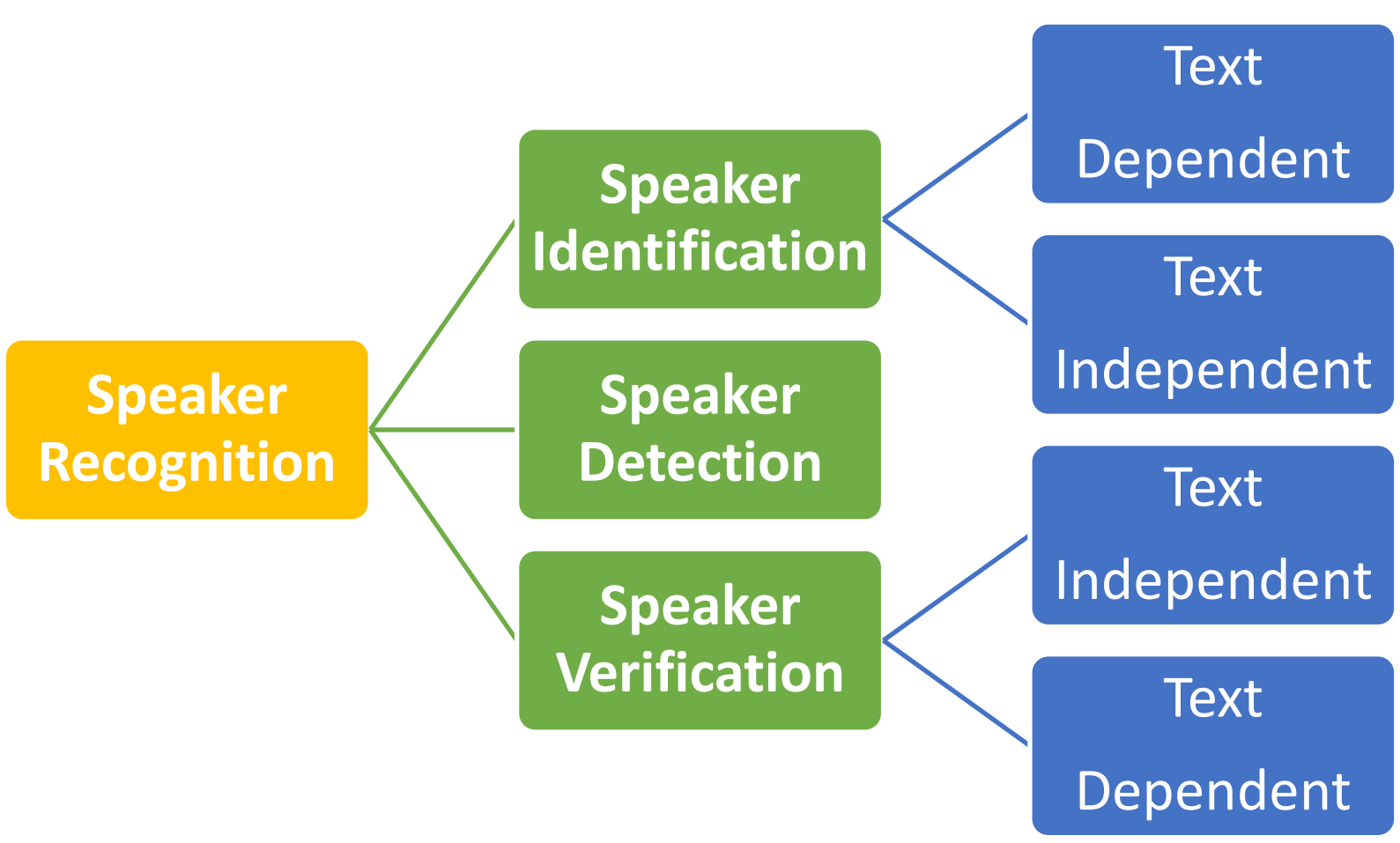

*Figure 2.The speaker recognition approaches*

Text dependency serves as an advanced classification within the realm of speaker recognition, hinging on the specific text articulated by the speaker during the recognition process. The two main subcategories of this categorization are text-dependent and text-independent recognition. Text-dependent automatic speaker recognition (ASR) refers to a method in which the spoken test utterance must align precisely with the text used in the enrollment phase (Heigold et al., 2016). Under this paradigm, the test speaker is familiar with the model already and the local lexicon runs under modest enrollment and trial phases in order to target exact outcomes. This method has some technical and scientific difficulties even if it has benefits. Key components supporting contemporary methodologies, like advanced feature extraction, speaker modeling, and likelihood ratio scoring score normalizing techniques, were revealed by the development of text-dependent speaker identification systems in the 1990s. Many architectures have been studied throughout years to improve this field of inquiry. Text-independent ASR, on the other hand, entails training and testing stages in which the speech signals remain completely free from control (Bimbot et al., 2004). Usually for both training and testing, this approach calls for lengthier speech segments; moreover, the test speaker is not familiar with the enrollment samples at all. Greater freedom provided by text-independent recognition lets the speaker engage with the system more organically than by its text-dependent equivalent. But with this method, reaching ideal accuracy calls for long training and plenty of testing. Further subcategories of both text-dependent and text-independent methods are open and closed set speaker recognition.

The open set method lets the system process speech from a wide range of unknown speakers and allows an arbitrary number of trained speakers. By comparison, the closed set technique is limited to a certain group of registered speakers, whereby the system seeks to detect a speaker's distinct vocal traits from a predetermined collection of recognized voices. Although this study is mostly focused on the approaches linked to speaker recognition, the domain of speech and speaker recognition also includes several ways for identifying and assessing the emotional content of speech. Particularly with regard to applications like voice command systems, speech technology, and speech analysis, the relevance of speaker identification and verification has greatly grown inside the research community. Researching speech characteristics and running searches depending on speaker identification is attracting more and more attention. This increased focus on speaker identification systems highlights the large corpus of studies produced over the years, which has been supplemented by multiple literature assessments looking at many aspects of speaker recognition.

Three main categories might help one classify the literature on speech recognition (SR). The first category consists of thorough studies examining the corpus of current knowledge on different kinds of SR and general SR processes; they will be discussed in the next part. Important works falling within this category are research (El Ayadi et al., 2011) (Lawson et al., 2011) (Saquib et al., 2010). The second category mostly investigates statistics and machine learning techniques used as SR classifier. Important contributions in this field consist in research (Farrell et al., 1994), (Larcher et al., 2014) (Lippmann, 1989) . This group strongly corresponds with the fields of categorization and machine learning, which generate a lot of research. Within the particular framework of speaker identification (SI), a quick study covering SI procedures and contrasts numerous general SI approaches (Sidorov et al., 2013). The final category is on SR's feature extracting techniques. One striking example is the most recent poll (Dişken et al., 2017), It explores techniques for obtaining strong speaker-specific characteristics considering several elements like noise patterns, emotional states, and channel mismatches. Besides Rao and Sarkar (Rao et al., 2014) gave a succinct summary of feature-based and model-based speaker verification methods. Chava and Chougule (Chavan and Chougule, 2012) also contributed a brief review that defines and explicates features relevant to speaker recognition. Moreover, Tirumala and Shahamiri (Tirumala and Shahamiri, 2016) underscored The importance of deep learning approaches for feature extraction in speaker identification. Further noteworthy contributions include the review by Jayanna and Prasanna (Jayanna and Prasanna, 2009), which discusses SR analysis, modeling, and feature extraction, as well as a survey by Lawson et al. [5] that evaluates acoustic features in SR through experimental assessments.

The experiment and progression of these systems have consistently shown a bias toward linguistically homogeneous environments, particularly in relation to dialect-rich languages such as English or Mandarin, where phonetic and lexical variations are minimal. This focus simplifies feature standardization and system design but limits applicability to languages that exhibit significant dialectal diversity, such as Kurdish. The underrepresentation of diverse languages highlights a critical gap that needs to be addressed to create technologies that are inclusive on a global scale.

## 3. Kurdish Language

This part looks at the several dialects of the Kurdish language, therefore offering a thorough study of the Central Kurdish writing system together with important pronunciation issues. Such understanding is essential as it shapes the automatic lexicon creation in our work. There are various dialects in the Kurdish language, three main groups may be distinguished from each other. In Iran and Iraq, most of the Kurds speak Central Kurdish, sometimes known as Sorani. The official written standard of Central Kurdish emerged in the 1920s. (Veisi et al., 2022) later came to be accepted as the official Kurdish orthography for Iraq. On the other hand, Northern Kurdish, often referred to as Kurmanji, is quite common in the northern parts of Kurdistan, including Turkey, Northern Iraq, Northwestern Iran, and Northern Syria. Kurds in the Kermanshah and Ilam provinces of Iran as well as in the southern areas of Iraqi Kurdistan, particularly in the Khanaqin district, make up the third group, Southern Kurdish.

Academic efforts by writers like Matras (Matras, 2017) explores these dialects more closely and notes several alternate categories. As a matter of fact, Ahmadi (Ahmadi, 2020) notes Laki as a possible fourth group among Kurdish speakers. Furthermore often regarded as different variants of Kurdish, Zazaki and Gorani fall to the Northwestern branch of the Iranian languages within the Indo-European family (Veisi et al., 2022). Northern Kurdish (Kurmanji) and Central Kurdish (Sorani) are two main dialects of Kurdish. Usually, Northern Kurdish is written in Latin (Yakubovskyi and Morozov) script, However, it could also be spelled in Arabic in Kurdistan Region of Iraq. In contrast, Central Kurdish is largely inscribed using a modified Arabic script. While Northern Kurdish boasts a larger speaker population, Central Kurdish has a richer repository of written materials.

Research on speech recognition in the Kurdish language remains limited, with only two studies currently published. The first of these studies was conducted by Abdullah and Veisi (Abdullah and Veisi, 2022). The study focuses on developing an ASR system for the Central Kurdish language (CKB) utilizing Deep Neural Network (DNN) transfer learning. The integration of Mel-Frequency Cepstral Coefficients (MFCCs) for speech signal feature extraction, alongside Long Short-Term Memory (LSTM) with a Connectionist Temporal Classification (CTC) output layer, is employed to develop an Acoustic Model (AM) utilizing the AsoSoft CKB speech dataset. Moreover, they employed the N-gram language model on the amassed extensive text dataset, comprising over 300 million tokens. The text corpus is utilized to construct a dynamic lexicon model comprising over 2.5 million CKB terms. The observed results demonstrate that using a DNN enhances results when compared to standard statistics modules. The proposed strategy, which combines transfer learning and language model modification, yields a 0.22%-word mistake rate. This outcome surpasses the most accurate performance recorded for the CKB. The second paper covers the Jira system (Veisi et al., 2022) . Originally intended especially for the CKB language, the Jira ASR system is the first sizable vocabulary voice recognition system (LVSR). Many conventional techniques—including HMM-based models and SGMM approaches—were used to build the acoustic model. Researchers developed a phrase collecting di-phone ratio that faithfully captures the subtleties of the CKB language in building the voice corpus. 576 people in all noted the desired words during a total 43.68-hour period. Two settings were used for these recordings: a social network context employing mobile phones (referred to as AsoSoft Speech-Crowdsourcing) and a controlled environment using a noise-free microphone (called AsoSoft Speech-Office). With an average word error rate of 13.9% throughout many document themes and an outstanding 4.9% for the total subject matter, the SGMM acoustic model produced the most significant results.

## 3.1 Challenges in Kurdish Speech Processing

For voice recognition research, the Kurdish language ecology provides a very sophisticated case study. Unlike many conventional techniques for language identification, Kurdish offers a multifarious terrain marked by notable dialectal variation. Linguists such as Hassanpour (Hassanpour, 1998) contend that these dialects reflect complicated communication systems with strong historical and cultural origins, not just variances of a language. Unique phonological, lexical, and grammatical characteristics of every dialect provide difficulties for traditional voice recognition systems. Technological methods of Kurdish speech identification have always had numerous major obstacles. Comprehensive research projects have been particularly hampered by limited computing resources, scarce linguistic documentation, and geographical fragmentation of Kurdish-speaking areas.

For voice recognition technology, the phonetic variation across Kurdish dialects presents difficult problems. Mostly spoken in northern Turkey, Syria, and areas of Iraq, Kurmanji has somewhat distinct phonological traits than Sorani, more often found in Iraq and Iran. Often regarded as a more specialized dialect, Hawrami adds even more intricacy with its own phonetic and grammatical patterns. Furthermore important for Kurdish voice recognition research is code-switching. The geographical setting of Kurdish-speaking areas—spanning many national borders—has promoted a linguistic environment marked by complex language exchanges. Frequently switching between Kurdish and surrounding languages—such as Turkish, Arabic, Persian, and Armenian— Kurdish speakers create a complex linguistic ecology that standard models of recognition find difficult to manage. Multiple levels of complexity are introduced by this code-switching phenomena, hence testing current machine learning methods meant for more homogenous language settings.

Kurdish voice recognition presents methodological difficulties beyond the technical ones. Development of good recognition systems depends critically on sociolinguistic elements. Crucially, multidisciplinary research combining computational linguistics, machine learning, and a strong cultural awareness calls for integration. Speaker identification technologies have to include not just auditory variances but also social, historical, and contextual language complexity. Promising answers to the difficulties of dialect detection are starting to come from technological developments in deep learning and neural network design. Modern study by Bahdanau et al. (2016) and later investigations by Mohamed et al. (2018) have shown how well recurrent neural networks and attention-based models may capture complex language variances. These methods provide a more flexible way for dialectal voice detection by using cutting-edge machine learning algorithms that can dynamically adjust to phonetic variances. Novel approaches to dialectal variances are being brought by using computational methods. Overcoming the difficulties related to limited training data has showed promise with respect to transfer learning and domain adaption techniques. Researchers such as Graves et al. (2013) have shown how these methods may efficiently extend acoustic models across several dialects, hence perhaps generating innovations in Kurdish speech detection systems.

In the realm of speaker identification research, the junction of technology advancement and linguistic variety offers great possibilities as well as major difficulties. Recent research underline the need of adaptive, context-aware recognition systems able to efficiently negotiate challenging language settings. Future advancements must concentrate on building thorough datasets, effective acoustic modeling, and machine learning algorithms able to capture the complex intricacies of Kurdish dialects, according to language researchers and computer scientists. Empirical data emphasizes the indispensible necessity for specific, dialect-aware recognition strategies. Comparatively to over 90% accuracy in more standardized language environments, a historic investigation by Xanozi et al. (2021) indicated that generic voice recognition models attained accuracy rates as low as 30–40% when applied to variances in Kurdish dialects. This notable performance difference emphasizes how urgently specialized recognition algorithms that can fit the complex phonetic and lexical characteristics of Kurdish speech are needed.

## 4. Methodology

The procedures of gathering and getting ready the data set needed for this investigation are described in this part. The data collecting starts with compiling recorded speech samples from three different Kurdish dialects. Pre-processing methods are used on raw audio data following the collecting phase to guarantee it is correctly prepared and ready for use into the intended model. The pre-processed data is then split in two: a training set and a testing set.

## 4.1 Data Collection

Since the quality and variety of the data greatly affect the results of the study, the process of gathering datasets is a fundamental element of this work. With a fair mix of male and female contributors, the dataset consists of voice samples from more than 10 people, ranging in participation from common residents, television presenters, and political leaders. To guarantee linguistic variation, each participant sent voice recordings in three Kurdish dialects: Sorani, Kurmanji, and Hawrami.

Direct contributions combined with publicly available resources helped us find the voice recordings. While some participants recorded and emailed their voices especially for this study, others' recordings came from websites like YouTube, including speeches or broadcasts by political leaders and TV personalities. Every recording ran longer than ten minutes and was converted to WAV to guarantee consistency and ease processing.

As Figure 3 offers a thorough analysis of the dataset's distribution throughout several categories. With 17% of the data, "Record Duration" is the biggest part shown in blue. "Record Sources: Public Platforms" and "Gender: Male," both with 14% and shown in dark yellow and light yellow correspondingly, come next. Smaller segments include "Record Sources: Specific Contributions" and "Gender: Female," at 9% each.

The "Participants' Position" category shows that "Ordinary Individuals" make up 8% of the data set (orange), while "TV Presenters" and "Political Figures" each account for 7% (purple and dark blue, respectively). Representing the three Kurdish dialects—Kurmanji (green), Sorani (dark blue), and Hawrami (brown)—each contribute 5%. The smallest categories each reflect Clear labels on a color-coded chart clearly show these groups together with their respective percentages. This exacting data collecting and classification guarantees a strong basis for study.

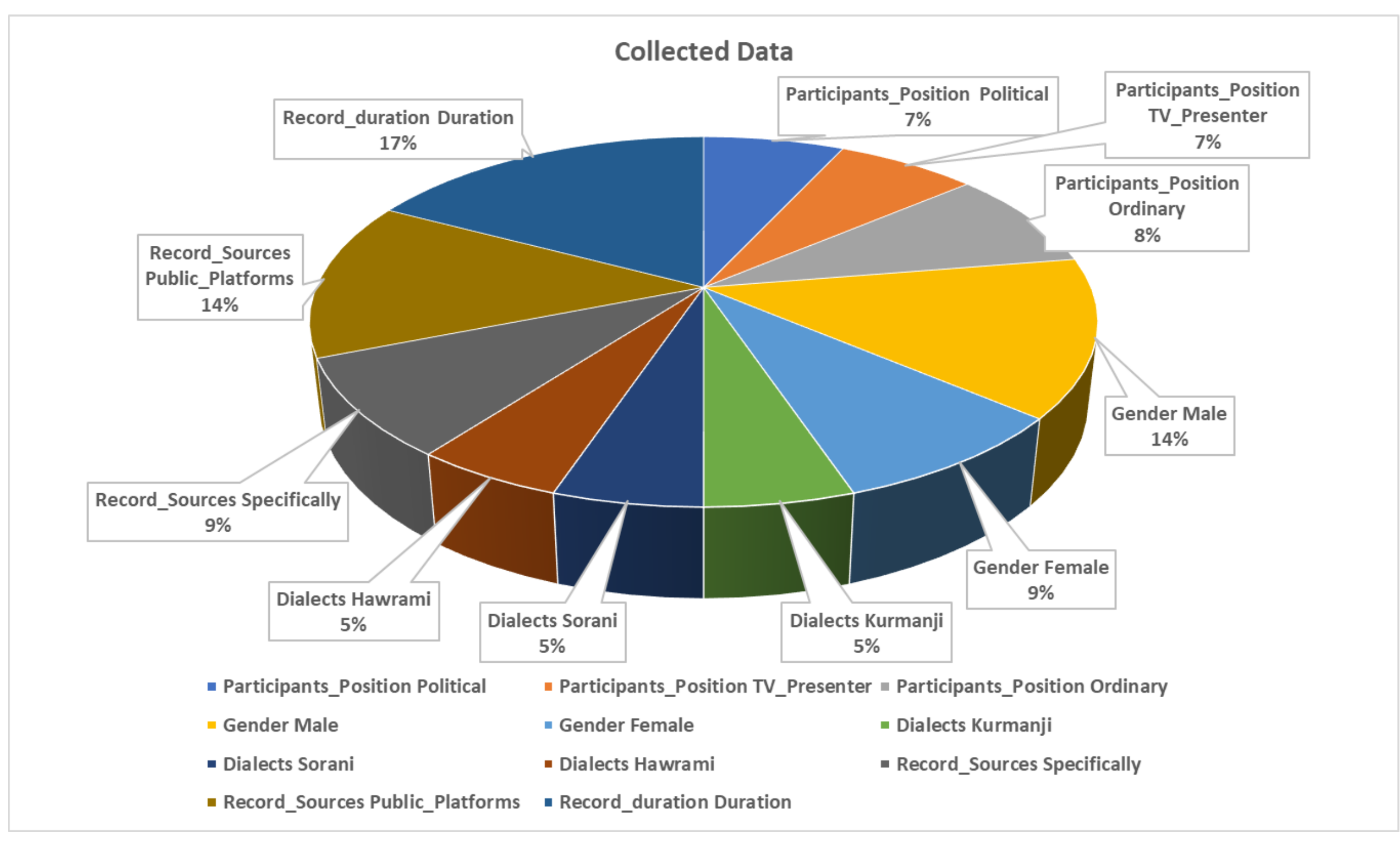

*Figure 3.Dataset Distribution*

## 4.2 Data Pre-processing

### 4.2.1 Splitting Data

Data collecting and pre-processing were followed by dataset preparation for use into the intended model. One had to split the data into separate subsets—a training set and a testing set—before using the processed data for model training and assessment. Every subset has a different function in the modeling process and will be discussed in their respective parts more especially.

#### 4.2.1.1: Training

The focus of this section is on the training data for the model, particularly using the **Sorani dialect** to evaluate its ability to recognize speakers. The attached chart titled **"Training Data"** visually presents the distribution of the dataset across multiple categories, highlighting the **Dialect Sorani** as the largest segment with **12 instances**, which emphasizes its central role in the training process. The **Record_Source Public_Source** and **Record_Duration Duration** each account for **10 instances**, showing their significant contribution to the data. Gender representation includes **7 instances** of **Gender Male** and **5 instances** of **Gender Female**. Smaller categories include **Record_Source Specifically** with **2 instances**, while participant positions are broken into **Political (3)**, **TV Presenter (2)**, and **Ordinary (4)**. By prioritizing the Sorani dialect in training, the model's performance will be tested to assess its ability to recognize speakers, even when exposed to different dialects. The chart uses distinct colors and callouts to clearly display the data balance and highlight the key focus areas within the training dataset which is shown in Figure 4.

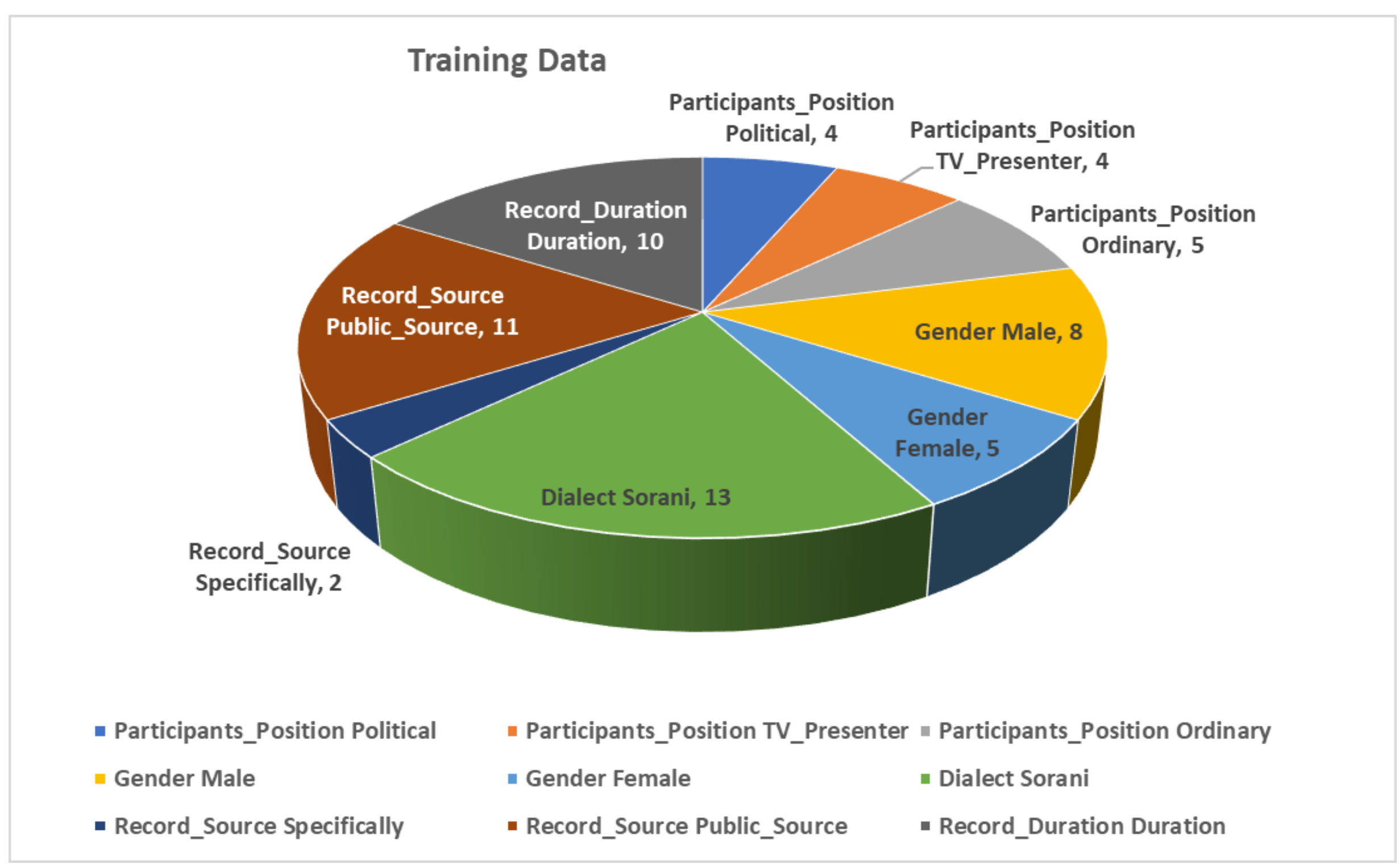

*Figure 4.Training Dataset*

#### 4.2.1.2 Testing

As shown in Figure 5 in this section, the focus will be on the **testing data** for the model, specifically evaluating its ability to recognize speakers using the **Kurmanji** and **Hawrami** dialects. The attached chart titled **"Testing Data"** visually illustrates the distribution of the dataset, with the **Speak Dialects Kurmanji** and **Speak Dialects Hawrami** categories being the largest, each accounting for **13 instances**. This highlights the emphasis on testing the model's performance with these two dialects to determine its robustness when exposed to different linguistic variations.

The chart also includes **Record_Sources Public_Platforms** with **11 instances**, indicating that a significant portion of the data comes from public sources, while **Record_Sources Specifically** contributes a smaller share with **2 instances**. Gender representation reveals five occurrences of gender female and eight instances of gender male. Reflecting a range of speaking duties, participant positions fall into Political (4), TV Presenter (4), and Ordinary (5) categories.

Examining the Kurmanji and Hawrami dialects in testing helps one to fully assess the model's generalizing and recognition of speakers across dialects. The testing dataset's makeup is clearly shown on the chart by means of different colors and callouts that highlight each group.

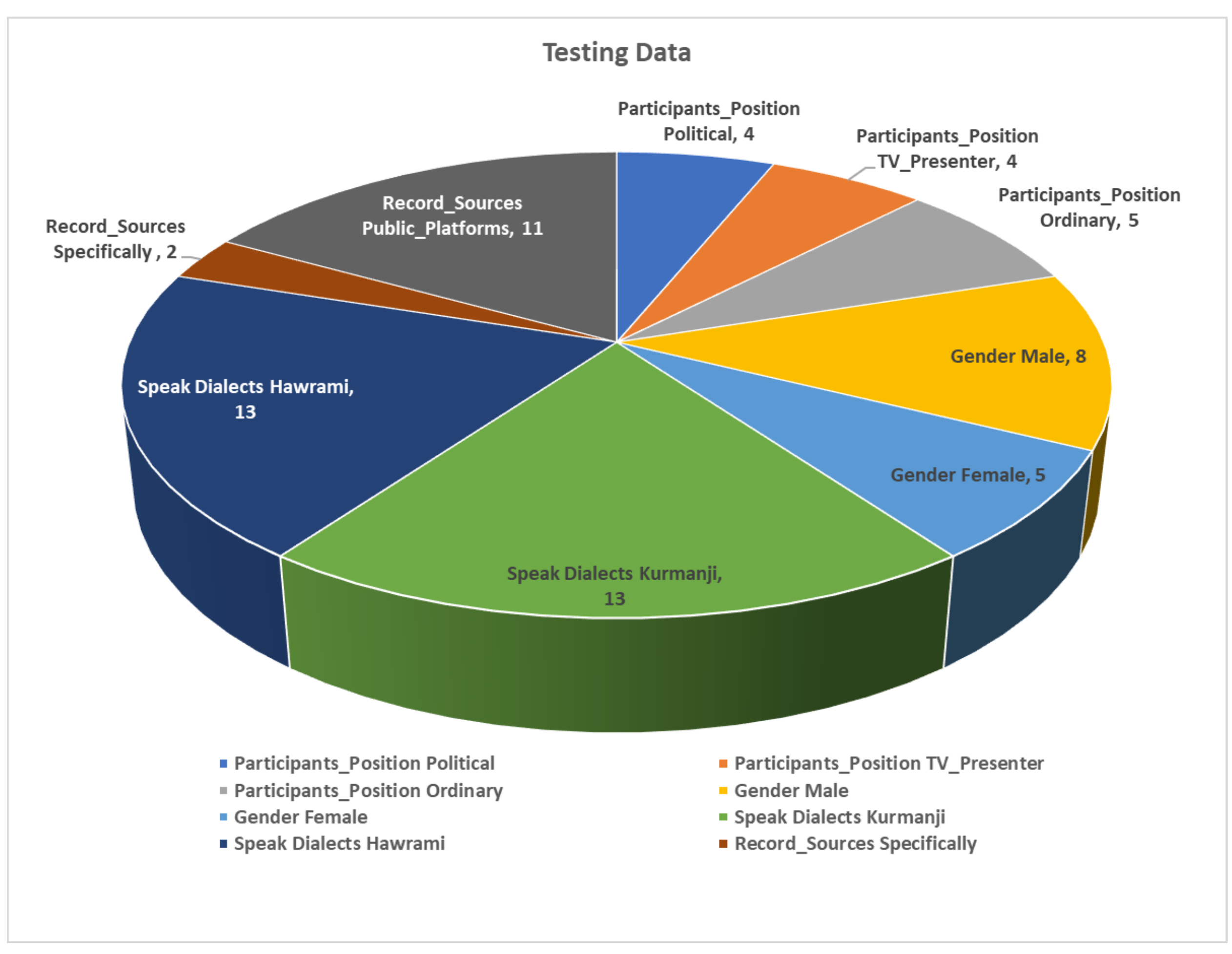

*Figure 5.Training Dataset*

## 5. Challenges and solution

### 5.1 Dialect Variation

MacKenzie (MacKenzie, 1961) among the first pioneering academics to classify Kurdish languages into "Northern" and "Central" groups, was Along with Southern Kurdish inclusion, this categorization has greatly affected Kurdish dialectology. Thus, although the names may vary, Kurdish dialects are usually categorized as Northern, Central, and Southern [2]. Spoken in Turkey, Syria, Iraq, and Iran, Kurmanji, sometimes Northern Kurdish, is a language.  Central Kurdish, often called as Sorani, is most found in Iran and Iraq.  Conversely, Southern Kurdish is primarily found in the Kermanshah and Ilam provinces of Iran, as well as in the surrounding regions of Iraq. (Eberhard et al., 2022). Languages such as Zazaki and Gorani are frequently perceived as distinct from Kurdish; however, they may alternatively be regarded as additional dialects. This perspective persists even though many speakers of these languages identify as ethnically Kurdish (Öpengin, 2024). The classification of Laki as the southernmost variant within Kurdish is still a matter of debate.  As clarified by Anonby, there is a consensus that Lori (Luri) is a Southwestern Iranic language, which challenges the notion that it is a variant of Kurdish. (Anonby, 2004). Nevertheless, Lori and Laki exhibit similarities with Southern Kurdish dialects. The Kurmanji speech zone is divided into five regions, while the Sorani continuum comprises two zones: Southern Sorani, centered in Sulaymaniyah, Iraq, and Northern Sorani, based in Erbil, Iraq.

It is important to mention some other details related to the recorded data which is presented in Table 1. General details about the data have been presented such as, total audio length include 130 hours and 35 seconds, all the audio files are stored in and used in (WAV) format file. Moreover, the longest time record is 10 Minutes and 26 seconds, however the shortest recorded sound is 7 minutes and 05 seconds, however only one audio file is that short the majority of the audio file recorded more than 10 minutes.

*Table 1. Recorded Data*

| Features | Details |
|---|---|
| **Total Audio Length** | 130 Hours 35 Seconds |
| **Audio File Format** | .wav |
| **Longest Audio File** | 10 Minutes and 26 Seconds |
| **Shortest Audio File** | 7 Minutes and 05 Seconds |
| **Average Audio Length** | 10 Minutes and 2.9 Seconds |

Clearly, these dialects present complex challenges for Kurdish speech recognition that require innovative and multifaceted solutions. This is because these dialects include profound phonetic and lexical differences across Kurdish linguistic regions. There are substantial acoustic variations among the Kurmanji, Sorani, and Hawrami dialects, emphasizing the intricate phonological distinctions that complicate traditional speech recognition methods. These differences are basic obstacles that call for complex computational methods, not only language curiosities.

Techniques of transfer learning offer a sophisticated approach to handle dialectal variants. Researchers can build more flexible and strong recognition systems by designing basic models that can dynamically change to fit various Kurdish language environments. Wang et al (Wang et al., 2021)  showed that speaker identification performance of dialect-specific models might be improved by well crafted transfer learning approaches. Future studies should concentrate on developing adaptive recognition algorithms able to traverse the complex terrain of Kurdish language. This work requires not just technological creativity but also a whole review of voice recognition techniques. Particularly those using attention processes and transfer learning, advanced neural network designs provide the most potential answers to the difficult problems of detecting Kurdish dialectal speech.

Advancement of effective voice recognition algorithms for the Kurdish language is much hampered by data shortage. Unlike more generally researched languages with rich linguistic resources, Kurdish dialects suffer a constant lack of complete speech datasets. Ahmad et al. (Ahmadi, 2020) shown that most current speech recognition methods rely on training sets either too small or insufficient in catching the minute dialect changes. Researchers have been looking at creative ideas to address this problem including synthetic data creation, crowd-sourced data collecting methods, and data augmentation approaches.

One very interesting approach to solve the constraints of current datasets is synthetic data creation. Modern generative models offer possible ways to increase constrained language datasets: voice conversion methods and generative adversarial networks (GANs). a trailblazing analysis by Hassan et al. (Hassan et al., 2017) shown that well-designed synthetic speech production may efficiently augment current training data, hence perhaps improving model resilience in dialectally varied settings. One developing answer to the problems of data shortage and language variety is crowdsourced data collecting. Platforms that enable distributed language data collecting have the ability to democratize research in voice recognition, hence producing more representative datasets.

## 5.2 Phonetic Similarities and Overlap

Developed in the 1920s, the Arabic-based writing system for Central Kurdish has changed many times since. Notably, this writing system is primarily phonemic, which means that each letter usually corresponds to a single phoneme, though there are some exceptions. For example, the letter 'ی' can represent both /j/ (a palatal approximant) and /i/ (a vowel). Similarly, the letter 'و' may be pronounced as /w/ (a bilabial approximant) or /ʊ/ (a vowel). When written as 'وو', it represents /u/ (a long vowel). Although this digraph functions as a single phoneme, it is not considered an individual character in the keyboard configurations for Kurdish that the Department of Information Technology of the Kurdistan Regional Government has released. (2014). Moreover, a short vowel is occasionally omitted in the Arabic script, as illustrated in the word "دڵ" [heart], where the phoneme /i/ appears between the letters "د" and "ڵ"; this phoneme is, however, represented in the Latin script of Kurdish.

The challenge posed by phonetic similarities and overlaps among Kurdish dialects presents a complex computational problem. Speakers from different dialectal regions often have similar acoustic characteristics, which render traditional speaker recognition techniques ineffective. The advanced feature extraction methods utilize deep neural network architectures to capture subtle linguistic nuances. These approaches employ multi-layer acoustic modeling and attention-based mechanisms to effectively distinguish between speakers with high levels of phonetic similarity.

# 6. Machine Learning models

### 6.1Multinomial Naïve Bayes (MNB)

The multinomial naive Bayes (MNB) algorithm is extensively utilized in the domain of text classification, hinging upon the likelihood that a term *t* is associated with a specific class c, as illustrated in Equation 1.

$$p(c,t) = \frac{p(c)*p(t,c)}{p(t)} \qquad \text{eq(1)}$$

In the representation provided in Equation 1The MNB algorithm determines the likelihood of a term t being associated with class c. The probability of class c's occurrence is denoted by p(c), while the joint probability of term t appearing within class c is denoted by $p(t,c)$. The overall probability of term t appearing within the entire corpus is reflected by $p(t)$. The implementation of this algorithm facilitates the feature selection process by evaluating the probability of term selection in relation to a specified class c (Bay and Çelebi, 2016).

### 6.2 Support Vector Machine

Support Vector Machines (SVM) are utilized to construct a model based on a training dataset, defined by feature vectors. By building a border separating positive from negative data, Support Vector Machine (SVM) classifies it. An efficient SVM classifier makes best use of the margin separating these classes. Once the model is built, it may be applied to evaluate fresh data and forecast corresponding categories.

$$W_x X_i + b \ \leq -1 \ \ For\ all\ X_i \ \epsilon \ \acute{c} \qquad \text{eq(2)}$$

$$W_x X_i + b \ \geq 1 \ \ For\ all\ X_i \ \epsilon \ c \qquad \text{eq(3)}$$

Where b denotes the bias value and w stands for the vector weights (Saeed et al., 2024).

### 6.3 K-Nearest Neighbor (KNN)

KNN is a categorization system. The basic idea of this identification method is to find the class of a given query depending not only on the closest text in the feature space but also on the classes of the k closest texts. It uses the Euclidean distance metric to classify works according to their subject matter into one or more predetermined categories. By means of a comparison with the keywords of fresh, unclassified texts, this classifier evaluates the keywords of existing texts. In particular, the KNN method finds the k keywords in the corpus most close to element y when one has a labeled corpus xi and wants to categorize a new element y. The categories related to these surrounding keywords then guide the categorization. The method consists on assessing possible text classifications by use of training examples with closest keywords to the text. The method sees related books as points in a Euclidean space where the Euclidean distance formula finds the distance between themes. Equation 3 allows one to get the distance for two points in this space, expressed as p = (x1,x2) and q = (a, b).

$$y = d(p,q) = d(q,p) = \sqrt{(x_1 - a)^2 + \ (x_1 - b)^2} \qquad \text{eq(4)}$$

The weighting of the classes represented by the surrounding texts is determined using the similarity score between each test text and every other text. When several k-nearest neighbors contribute to a given class, the weights linked to that class are aggregated and provide a cumulative weighted score that acts as the likelihood estimate for possible classes. This technique produces the test text's ordered ranking. Consequently, the categorization of text is conducted based on a thresholding method applied to these scores, ultimately leading to the identification of dual classes, as illustrated in Equation 4.

$$s(x,y) = \frac{x^t}{||x||\ ||y||} \qquad \text{eq(5)}$$

An equation such as $x^t$ can be defined as the transposition vector $x$, $||x||$ and $||y||$ can be defined as their Euclidean norm $y$, respectively (Kadhim, 2019).

## 6.4 Random Forst (Abderehman et al.)

The random forest Figure 6 recognized as one of the premier methods in machine learning. While it can also serve as a regression technique, its primary application lies in classification due to its versatility and user-friendliness. This approach combines multiple learning models, resulting in enhanced overall performance. By integrating a variety of uncorrelated trees, a random forest can achieve higher accuracy. Additionally, random forests can effectively impute missing values. Moreover, decision tree classifiers are celebrated for their remarkable performance; since random forests consist of a set of decision trees, they inherently exhibit greater robustness and power. A basic decision tree, when applied to classification challenges, can deliver impressive results with high accuracy (Hassan et al., 2022).

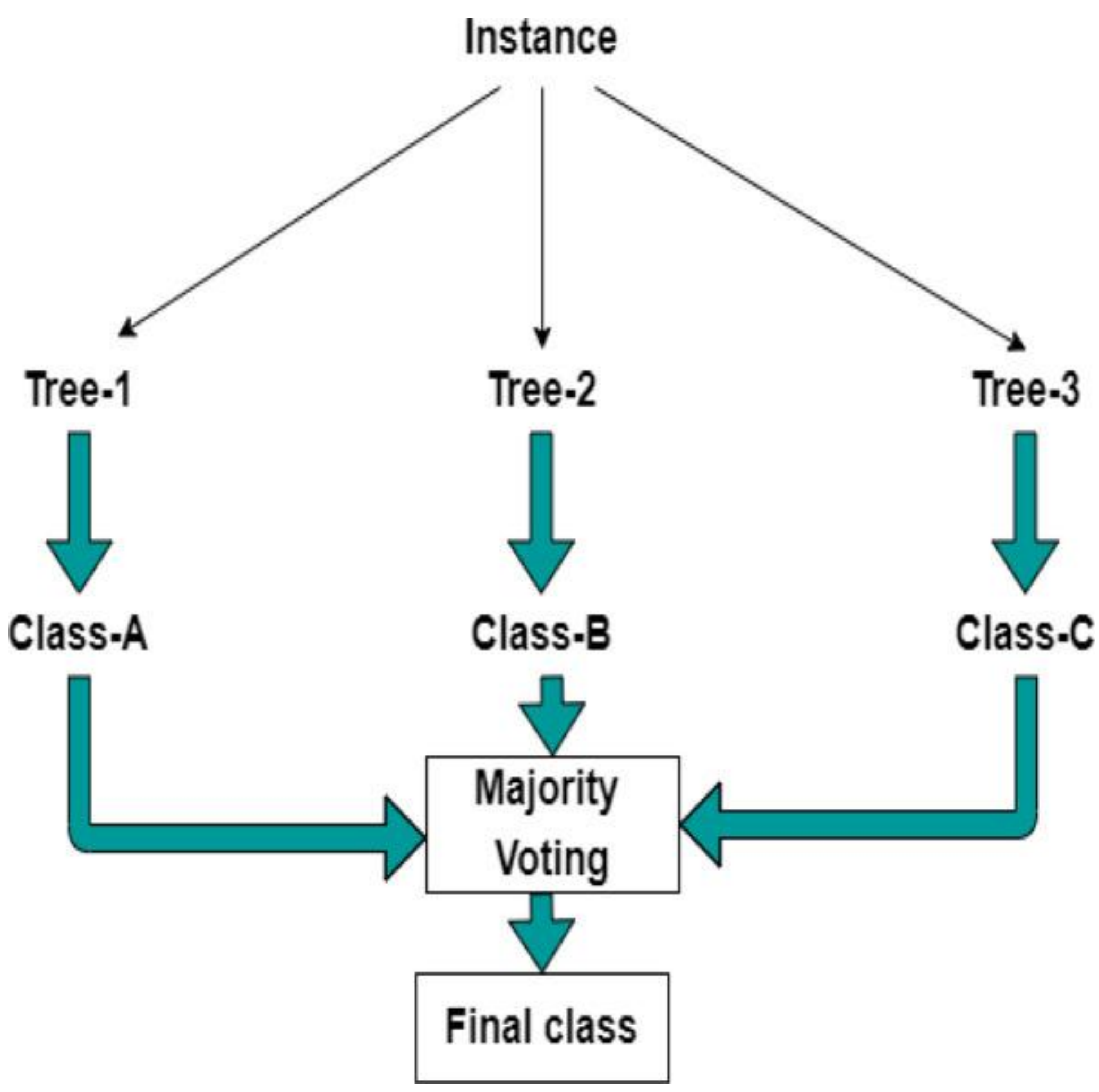

*Figure 6.Random Forest Classification*

## 6.5 Convolutional Neural Network (CNN)

This CNN architecture, designed for recognizing Kurdish speakers, features a carefully constructed model optimized for identifying distinct vocal patterns, as shown in Figure 7. The model begins with an initial convolutional layer (conv2d) that serves as the first stage of feature extraction. It processes what is likely a time-frequency representation of Kurdish speech, most probably a spectrogram or Mel spectrogram where time spans approximately 498 frames and frequency components are spread across 11 bands. This layer creates 32 unique feature maps with 320 learnable parameters, effectively detecting fundamental acoustic patterns such as phonetic components and tonal variations characteristic of Kurdish speech. Following this initial extraction, the architecture incorporates a max_pooling2d layer that reduces the time dimension by half and significantly compresses the frequency dimension. This results in a representation of dimensions (5, 249, 32) without adding any parameters. This strategic down-sampling preserves essential features while reducing computational demands and providing invariance to minor shifts in the input. Additionally, a dropout layer at the same dimensions introduces regularization by randomly deactivating neurons during training, based on a probability defined by the model's hyperparameters. This forces the network to develop redundant pathways for classification and prevents it from over-relying on specific features.

The second convolutional stage (conv2d_1) expands the feature depth to 64 channels while further reducing spatial dimensions to (3, 247, 64), using 18,496 parameters to detect more complex and abstract acoustic patterns. This layer likely captures speaker-specific characteristics such as vocal tract configurations, articulation styles, and dialectal nuances of Kurdish speakers. Another max_pooling2d_1 and dropout_1 sequence compress the representation further to dimensions (1, 123, 64), retaining only the most discriminative features. The classification pathway begins with a flatten operation that transforms the multi-dimensional feature maps into a single 7,872-dimensional vector, creating a comprehensive representation of the speaker's vocal characteristics. This flattened representation is then fed into a dense layer with 128 neurons, which contains the majority of the model's parameters (1,007,744). This layer performs the essential dimensionality reduction that maps the high-dimensional acoustic features to a more manageable latent space, making speaker differentiation more tractable. A dropout_2 layer prevents overfitting by randomly nullifying 128-dimensional vectors during training.

The final classification layer (dense_1) has 14 output neurons, indicating that this model is specifically designed to identify 14 distinct Kurdish speakers or potentially 14 different Kurdish dialects or accent variations. With just 1,806 parameters, this layer makes the final classification decision based on the learned representation. Overall, the entire model contains 1,028,366 trainable parameters (approximately 3.92 MB in storage), making it relatively lightweight and potentially deployable on edge devices or in resource-constrained environments, while still maintaining sufficient complexity to capture the nuanced vocal characteristics necessary for accurate Kurdish speaker recognition.

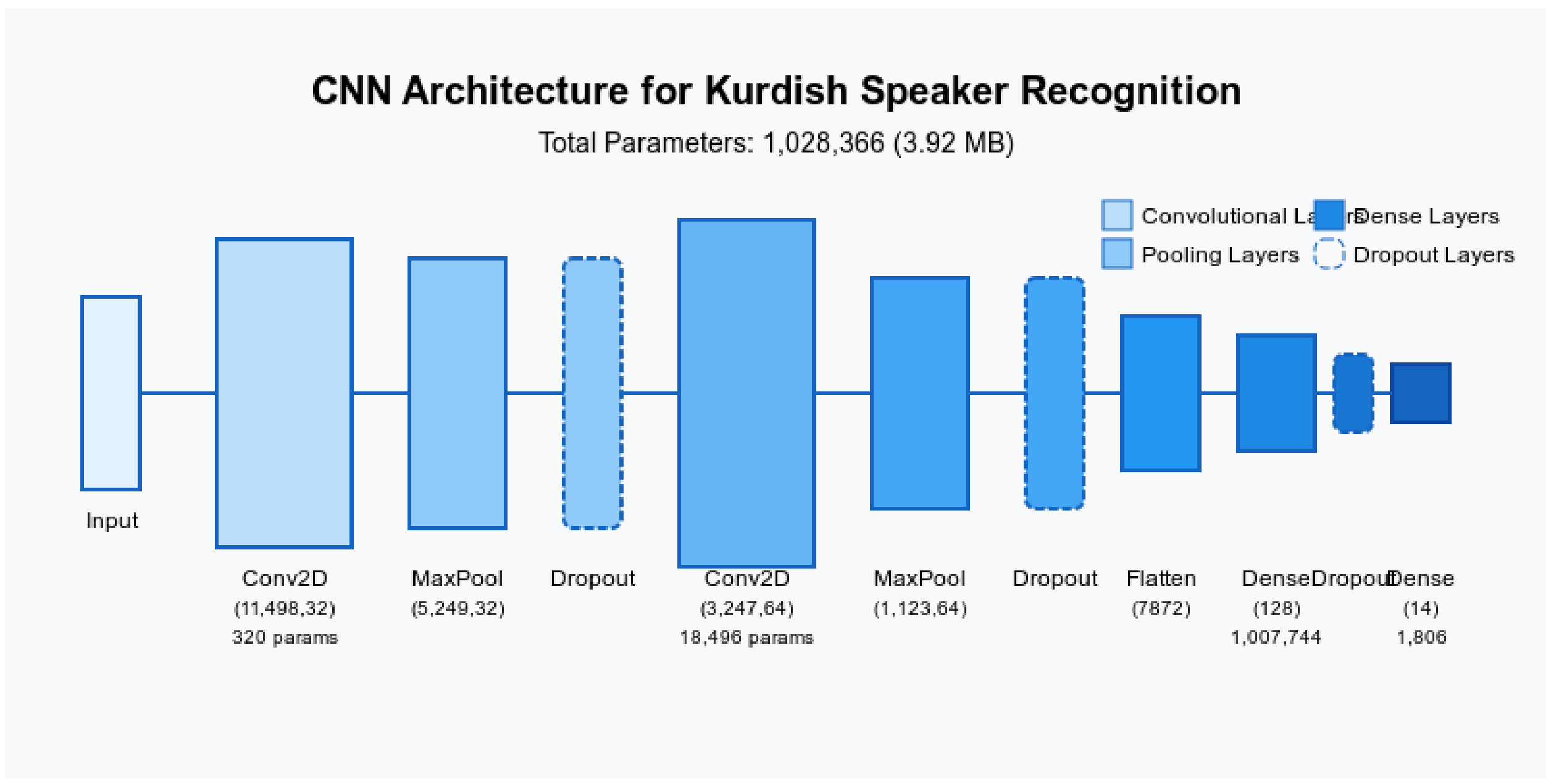

*Figure 7.The CNN architecture*

**7. Result and Discussion**

The field of Kurdish speech recognition represents an important area in computational linguistics, presenting researchers with significant linguistic challenges. Kurdish is not a single, uniform language; rather, it consists of a rich tapestry of dialects that complicate conventional speech recognition methods. This linguistic diversity spans several geopolitical regions, including Turkey, Syria, Iraq, and Iran, creating a unique challenge that goes beyond traditional language processing approaches.

Therefore, this study utilizes and analyzes the results of five machine learning models—SVM, CNN, KNN, RF, and MNB—applied to three primary Kurdish dialects: Sorani, Hawrami, and Kurmanji. The research evaluates these models based on various performance metrics, including F1-Score, Accuracy, Recall, and Precision, to assess their effectiveness in addressing the complexities of Kurdish speech recognition, as shown in Table 2.

*Table 2.The performance of each model on the dataset*

| **Model** | **Metric** | **Sorani** | **Hawrami** | **Kurmanji** |
|---|---|---|---|---|
| **SVM** | **F1-Score** | **0.98** | **0.61** | **0.80** |
| | **Accuracy** | **0.97** | **0.59** | **0.78** |
| | **Recall** | **0.97** | **0.55** | **0.75** |
| | **Precision** | **0.98** | **0.63** | **0.82** |
| **KNN** | F1-Score | 0.92 | 0.45 | 0.68 |
| | Accuracy | 0.90 | 0.43 | 0.65 |
| | Recall | 0.91 | 0.40 | 0.62 |
| | Precision | 0.93 | 0.48 | 0.70 |
| **CNN** | F1-Score | 0.95 | 0.52 | 0.73 |
| | Accuracy | 0.94 | 0.50 | 0.71 |
| | Recall | 0.93 | 0.48 | 0.69 |
| | Precision | 0.96 | 0.55 | 0.75 |
| **MNB** | F1-Score | 0.85 | 0.35 | 0.60 |
| | Accuracy | 0.83 | 0.32 | 0.57 |
| | Recall | 0.82 | 0.30 | 0.55 |
| | Precision | 0.87 | 0.38 | 0.63 |
| **RF** | F1-Score | 0.93 | 0.50 | 0.70 |
| | Accuracy | 0.92 | 0.48 | 0.68 |
| | Recall | 0.91 | 0.45 | 0.65 |
| | Precision | 0.94 | 0.53 | 0.72 |

The performance of speech recognition models across Kurdish dialects reveals a complex technological landscape that reflects significant linguistic challenges in Table 2. The SVM model stands out as the most robust approach, showing remarkable accuracy across various dialects. For the Sorani dialect, the SVM model achieves nearly perfect results, with F1 scores and accuracy rates consistently above 0.97. This exceptional performance highlights the model's effectiveness in capturing the linguistic nuances of this well-documented Kurdish dialect. Still, the striking differences in performance among dialects highlight the basic difficulties in Kurdish speech processing. The main challenge is the Hawrami dialect as all models fight to get significant recognition rates. Though the SVM model performs the best, it only gets a 0.61 F1 score for Hawrami, but Sorani gets an amazing 0.98. This significant difference validates the original research's focus on the intricate linguistic ecology of Kurdish, where dialects are not only variants but rather different communication systems with distinctive phonological traits.

Conversely, the CNN model offers a fascinating argument for adaptive voice recognition systems. Although it does not exactly match the performance of the SVM, the CNN shows a more constant performance across dialects. For Sorani, it gets a 0.95 F1; Kurmanji gets a 0.73; and Hawrami gets a 0.52. This slow down in performance fits the findings of the research on the difficulties caused by dialectal variances, implying that neural network designs might provide more flexible methods to capture these linguistic subtleties. The least successful method, the MNB model performs lowest across all dialects. Clearly showing the limits of conventional machine learning techniques when confronted with the complexity of Kurdish language variation, its F1 ratings range from 0.85 for Sorani to just 0.35 for Hawrami. This result greatly supports the demand of the original research for more advanced, context-aware recognition algorithms able to negotiate the complex language environment of Kurdish-speaking areas.

Empirically, the models' performance emphasizes how much specialized, dialect-aware recognition techniques are absolutely necessary. The extreme discrepancies in identification rates between Sorani and Hawrami dialects highlight the critical need for sophisticated computing methods able to accommodate the vast phonetic and lexical variances within the Kurdish language. According to the studies, there are various important directions for next progress. Promising answers to the problems of limited training data and dialectal variability are transfer learning and domain adaptation approaches. The differences in performance imply that, especially for underrepresented dialects such as Hawrami, future study should concentrate on creating more complete datasets. Moreover, the results underline the need of an interdisciplinary approach combining modern machine learning methods, computational linguistics, and a strong awareness of language variants based on deep culture. These findings have tremendous consequences going beyond just technical ones. Reflecting the larger sociolinguistic complexity of Kurdish-speaking areas marked by geopolitical fragmentation and complicated language relations, they The difficulties in voice recognition reflect more general problems of maintaining and comprehending language variety in this multifarious cultural terrain. The objective remains not just to get accurate speech recognition but also to adopt a sophisticated, polite approach that captures the rich linguistic legacy of Kurdish-speaking populations as technology develops. Figure 8 shows the performance differences particular to every accent. With an average F1 score of 0.97, Sorani shines out; Kurmanji comes second at 0.77 and Hawrami trail far at 0.58. This graphic emphasizes the linguistic complexity in processing Kurdish speech and numerically demonstrates the great difficulties in cross-dialect speaker detection.

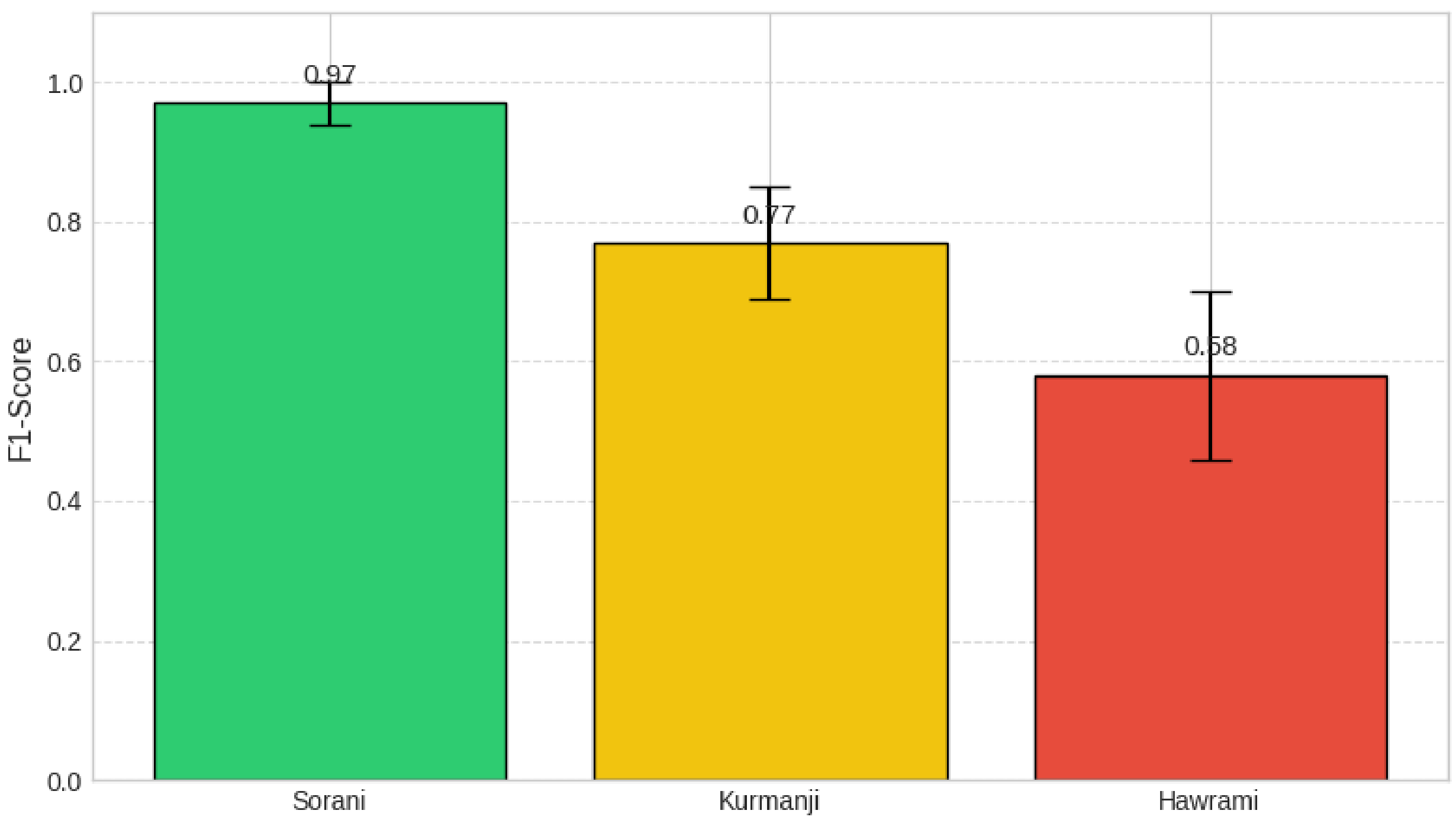

Figure 8. Mean F1-score

After the SVM model's impressive performance in first evaluations of certain dialects, we performed a rigorous cross-valuation research to fully evaluate the generalizability and robustness of the model as given in Table 3, 4, 5. The remarkable near-perfect identification rates attained in the Sorani dialect inspired us to look at the efficacy of the model throughout other Kurdish dialects.

*Table 3.Kurmanji Dialect Test Results*

| Speaker | Precision | Recall | F1-Score | Support |
|---|---|---|---|---|
| Masroor Barzani | 0.85 | 0.78 | 0.81 | 113 |
| Nechirvan Barzani | 0.88 | 0.82 | 0.85 | 115 |
| Kamal Gwlpi | 0.79 | 0.75 | 0.77 | 55 |
| Abduladim Hawrami | 0.91 | 0.84 | 0.87 | 25 |
| Gulstan MohammedAmin | 0.72 | 0.68 | 0.70 | 41 |
| Dnya Majid | 0.80 | 0.75 | 0.77 | 30 |
| Masoud Barzani | 0.89 | 0.85 | 0.87 | 62 |
| Avin Aso | 0.75 | 0.70 | 0.72 | 34 |
| Zhilya Ali | 0.65 | 0.50 | 0.56 | 5 |
| Zhwan Qaradaxi | 0.83 | 0.80 | 0.81 | 10 |
| Shahyan Tahseen | 0.78 | 0.74 | 0.76 | 49 |
| Latif Nerwaei | 0.82 | 0.78 | 0.80 | 18 |
| Sanh Shareef | 0.77 | 0.73 | 0.75 | 15 |
| Avg. (Kurmanji) | 0.80 | 0.75 | 0.77 | 572 |

*Table 4.Kurmanji Dialect Test Results*

| Speaker | Precision | Recall | F1-Score | Support |
|---|---|---|---|---|
| Masroor Barzani | 0.65 | 0.60 | 0.62 | 113 |
| Nechirvan Barzani | 0.70 | 0.65 | 0.67 | 115 |
| Kamal Gwlpi | 0.58 | 0.50 | 0.54 | 55 |

| Abduladim Hawrami | 0.75 | 0.70 | 0.72 | 25 |
|---|---|---|---|---|
| Gulstan MohammedAmin | 0.50 | 0.45 | 0.47 | 41 |
| Dnya Majid | 0.60 | 0.55 | 0.57 | 30 |
| Masoud Barzani | 0.68 | 0.62 | 0.65 | 62 |
| Avin Aso | 0.55 | 0.50 | 0.52 | 34 |
| Zhilya Ali | 0.40 | 0.30 | 0.34 | 5 |
| Zhwan Qaradaxi | 0.62 | 0.60 | 0.61 | 10 |
| Shahyan Tahseen | 0.59 | 0.53 | 0.56 | 49 |
| Latif Nerwaei | 0.64 | 0.58 | 0.61 | 18 |
| Sanh Shareef | 0.57 | 0.50 | 0.53 | 15 |
| Avg. (Hawrami) | 0.61 | 0.55 | 0.58 | 572 |

*Table 5.Results for Sorani Dialect (Baseline)*

| Speaker | Precision | Recall | F1-Score | Support |
|---|---|---|---|---|
| Masroor Barzani | 1.00 | 1.00 | 1.00 | 113 |
| Nechirvan Barzani | 1.00 | 1.00 | 1.00 | 115 |
| Kamal Gwlpi | 1.00 | 1.00 | 1.00 | 55 |
| Abduladim Hawrami | 1.00 | 1.00 | 1.00 | 25 |
| Gulstan MohammedAmin | 0.95 | 0.95 | 0.95 | 41 |
| Dnya Majid | 1.00 | 1.00 | 1.00 | 30 |
| Masoud Barzani | 1.00 | 1.00 | 1.00 | 62 |
| Avin Aso | 0.94 | 1.00 | 0.97 | 34 |
| Zhilya Ali | 1.00 | 0.60 | 0.75 | 5 |
| Zhwan Qaradaxi | 1.00 | 1.00 | 1.00 | 10 |
| Shahyan Tahseen | 0.96 | 0.96 | 0.96 | 49 |
| Latif Nerwaei | 0.95 | 1.00 | 0.97 | 18 |
| Sanh Shareef | 1.00 | 0.93 | 0.97 | 15 |
| Avg. (Sorani) | 0.98 | 0.96 | 0.97 | 572 |

Important new understanding of the linguistic complexity and technology difficulties related with Kurdish voice processing is revealed by the cross-dialect evaluation of the SVM speaker recognition model. Fundamentally, the results draw attention to notable variation in identification performance among many Kurdish languages. Notably, there is a stark contrast between the nearly perfect recognition rates in the Sorani dialect and the noticeably reduced accuracy in Kurmanji and Hawrami dialects.

The Sorani dialect is the criterion for outstanding performance as demonstrated in Figure 9, Having an average F1-score of 0.97, speaker recognition is really near-perfect. With an F1-score of 1.00 most speakers in this dialect show that the SVM model has been successfully trained and calibrated for this particular language variant. This outstanding performance emphasizes the best possible performance of the model in its most known dialectal environment. With an average F1-score of 0.77, the Kurmanji dialect shows a notable drop in recognition ability instead. While this still reflects a reasonable level of accuracy, certain speakers—such as Abduladim Hawrami, Nechirvan Barzani, and Masoud Barzani—consistently achieve higher recognition rates, suggesting that individual speech characteristics can sometimes transcend dialectal boundaries. Nevertheless, performance variation is considerable, as evidenced by speakers like Zhilya Ali, who experience recognition accuracy as low as 0.56, illustrating the significant challenges involved in cross-dialect speaker identification. The Hawrami dialect presents the most considerable obstacles for speaker recognition, with the SVM model achieving a significantly lower average F1-score of 0.58. This notable performance drop underscores the linguistic complexities that render Hawrami particularly difficult to process. Although top performers like Abduladim Hawrami, Nechirvan Barzani, and Masoud Barzani maintain recognition rates of around 0.65-0.72, the overall performance indicates substantial challenges in applying speaker recognition technologies across diverse dialectal contexts.

*Figure 9.Rader Chart on each dialect*

An intriguing observation is the consistent performance of certain speakers across dialects. Nechirvan Barzani, Masoud Barzani, and Abduladim Hawrami demonstrate relatively stable recognition rates, suggesting that some individual speech characteristics may have a degree of transferability beyond strict dialectal boundaries. Zhilya Ali, on the other hand, constantly shows the lowest performance across all dialects, suggesting that recognition accuracy is much influenced by personal speech patterns.

These results have significant consequences for speech recognition, particularly in linguistically varied environments. The findings confirm past studies emphasizing the complex linguistic ecology of Kurdish, where dialects serve not just as variants but also as separate communication systems with different phonological traits. The huge variances demand a more complex method to voice recognition, one that can dynamically adjust to language differences by aggregating powerful machine learning techniques with a rich cultural and linguistic knowledge. In the end, the research emphasizes how urgently more sophisticated computing methods are needed. Approaches include domain adaptation, transfer learning, and adaptive modeling are very vital for tackling the difficulties of cross-dialect speaker recognition. More thorough datasets, improved acoustic modeling methods, and identification algorithms able to negotiate the rich and complicated terrain of Kurdish language variation should be the main priorities of future study. This study is not just a technological challenge but also an important endeavor to protect and comprehend the complex language legacy of Kurdish-speaking populations. Figure 10 shows for each of the three Kurdish dialects distinct speaker recognition performance. With most speakers obtaining almost flawless F1 scores—that is, close to 1.00—the graph reveals a startling constancy in the Sorani dialect. By comparison, the Kurmanji and Hawrami dialects show great variation; F1 values range from 0.34 to 0.87. Notable constant actors that keep rather good identification rates throughout dialects include Nechirvan Barzani and Masoud Barzani.

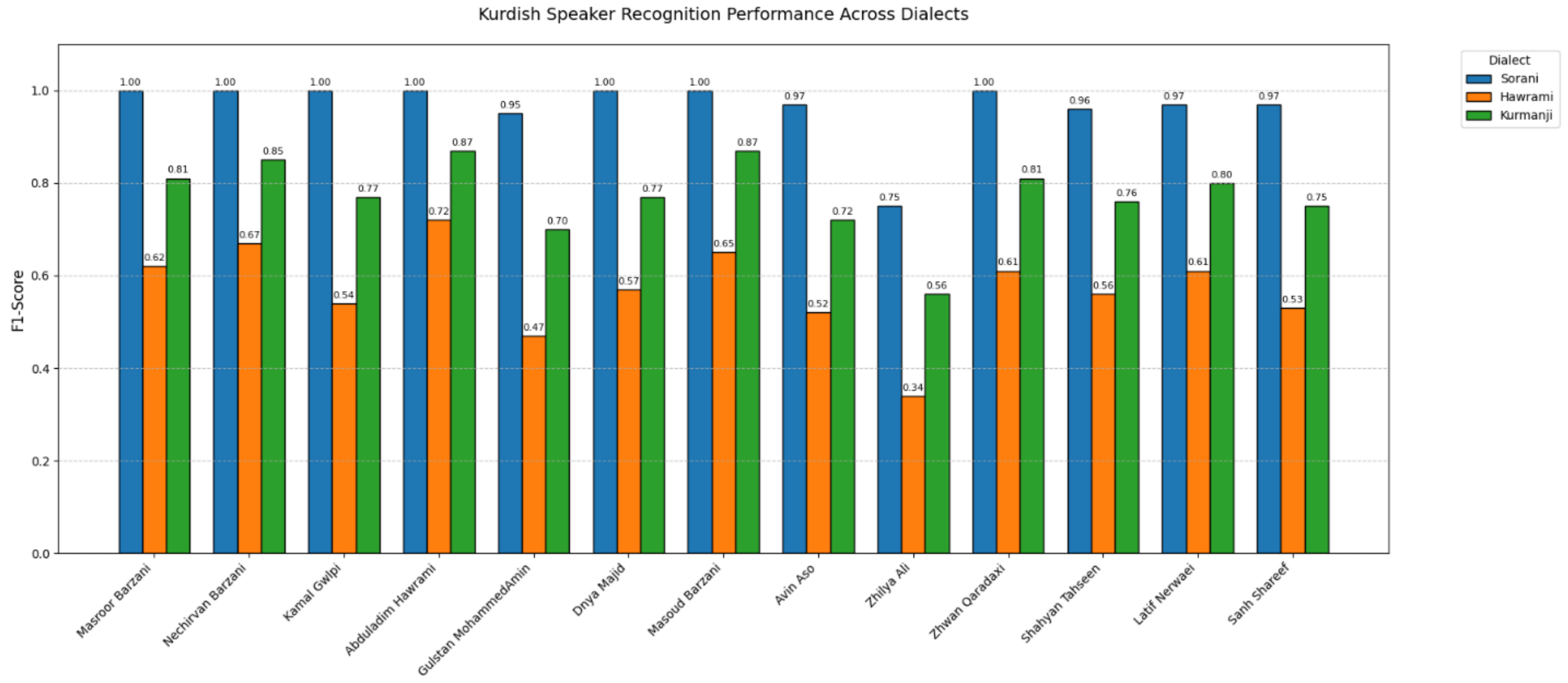

*Figure 10.Cross-Dialect Performance Comparison (Grouped Bars)*

Figure 11 provides a nuanced view of model performance when trained on one dialect and tested in another. The diagonal represents baseline performance (Sorani: 0.97, Kurmanji: 0.77, Hawrami: 0.58). The off-diagonal values reveal minimal cross-dialect transferability, with recognition scores dropping to as low as 0.08-0.23 when models are applied to dialects outside their training context.

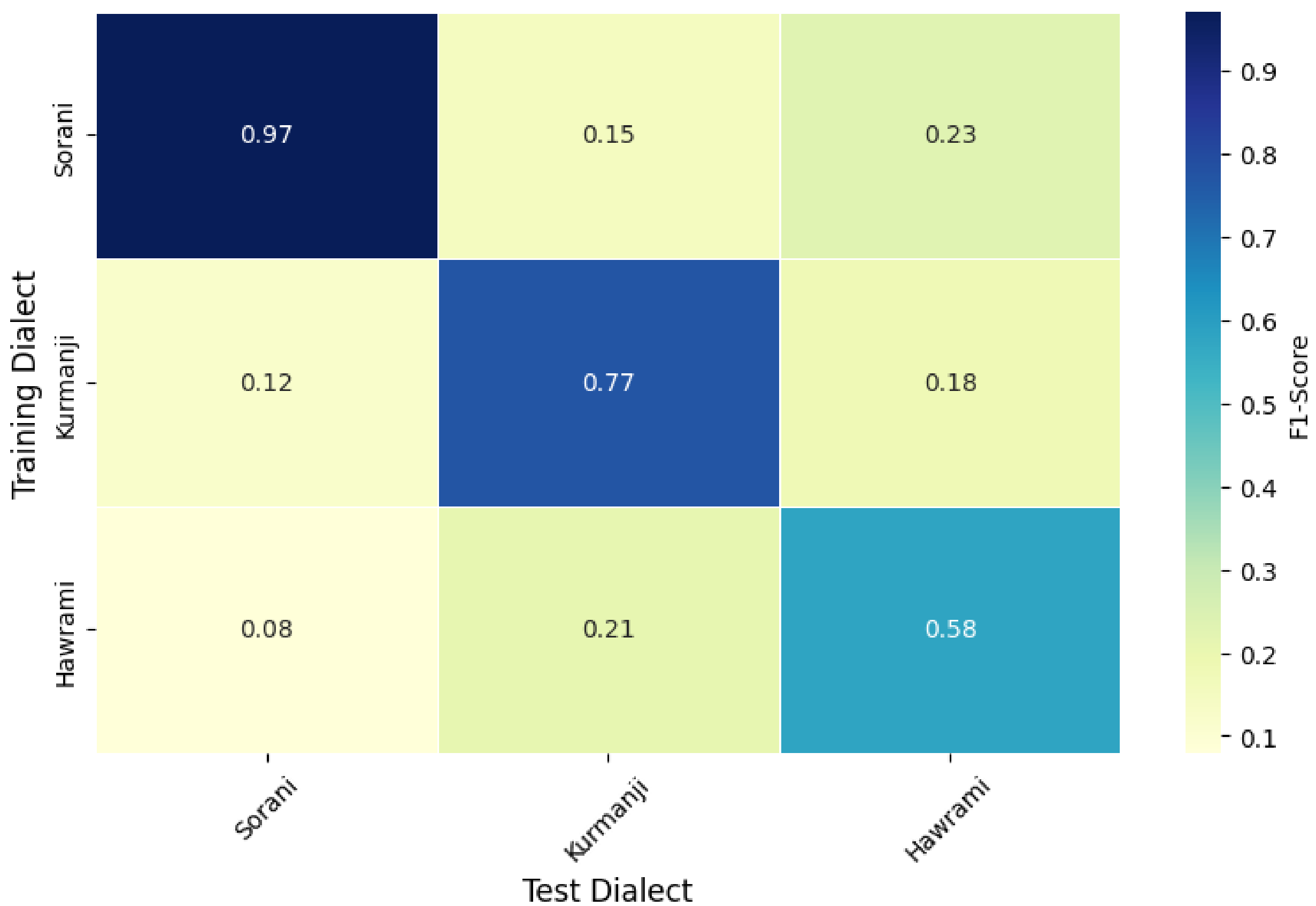

*Figure 11.Cross-Dialect Recognition Performance Heatmap*

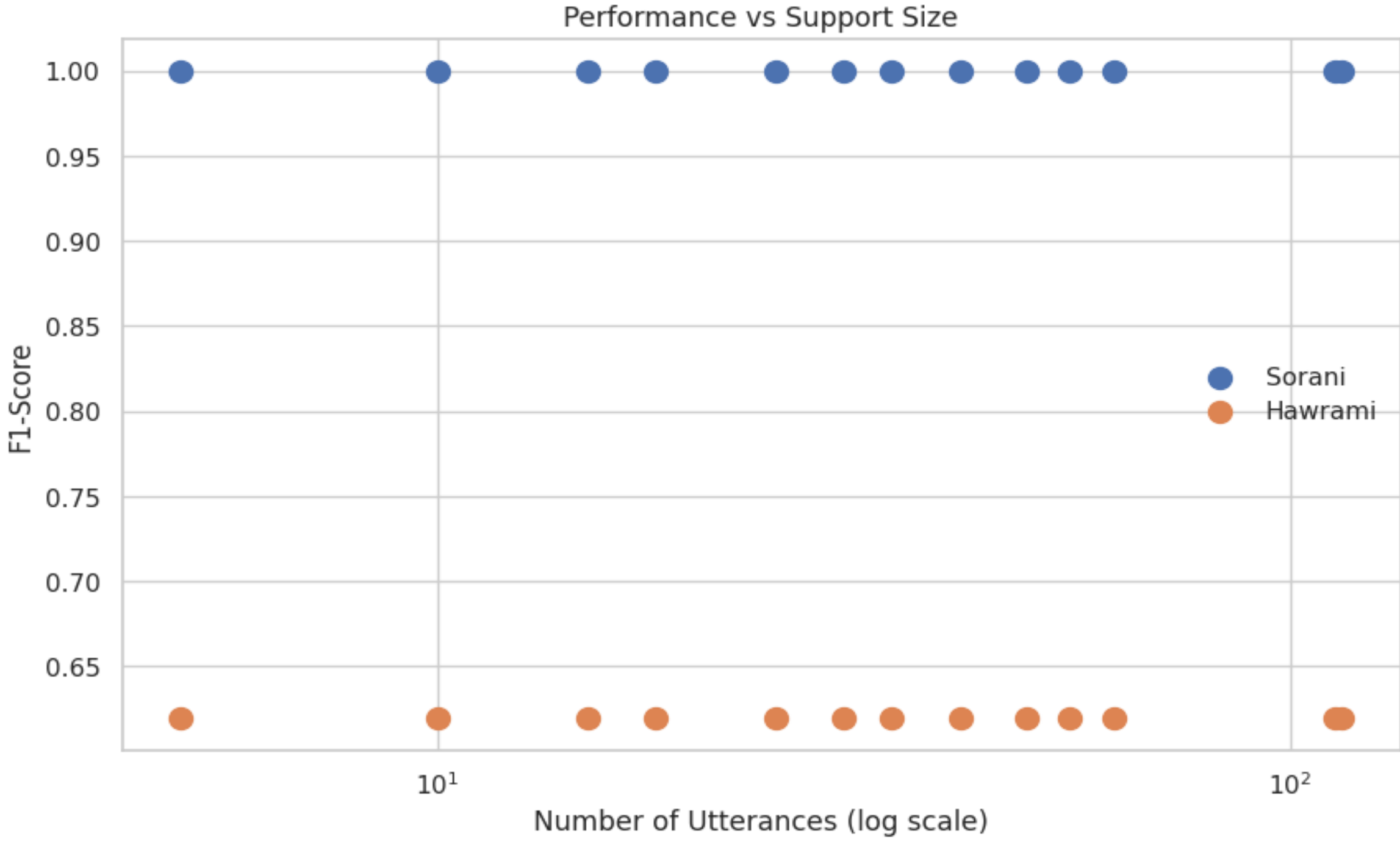

*Figure 12.Dialect Robustness*

Figure 12 provides a comprehensive visualization of dialect robustness, illustrating the relative performance across Sorani, Kurmanji, and Hawrami dialects. The asymmetric shape emphasizes the significant variations in recognition capabilities, with Sorani clearly dominating in terms of model performance.

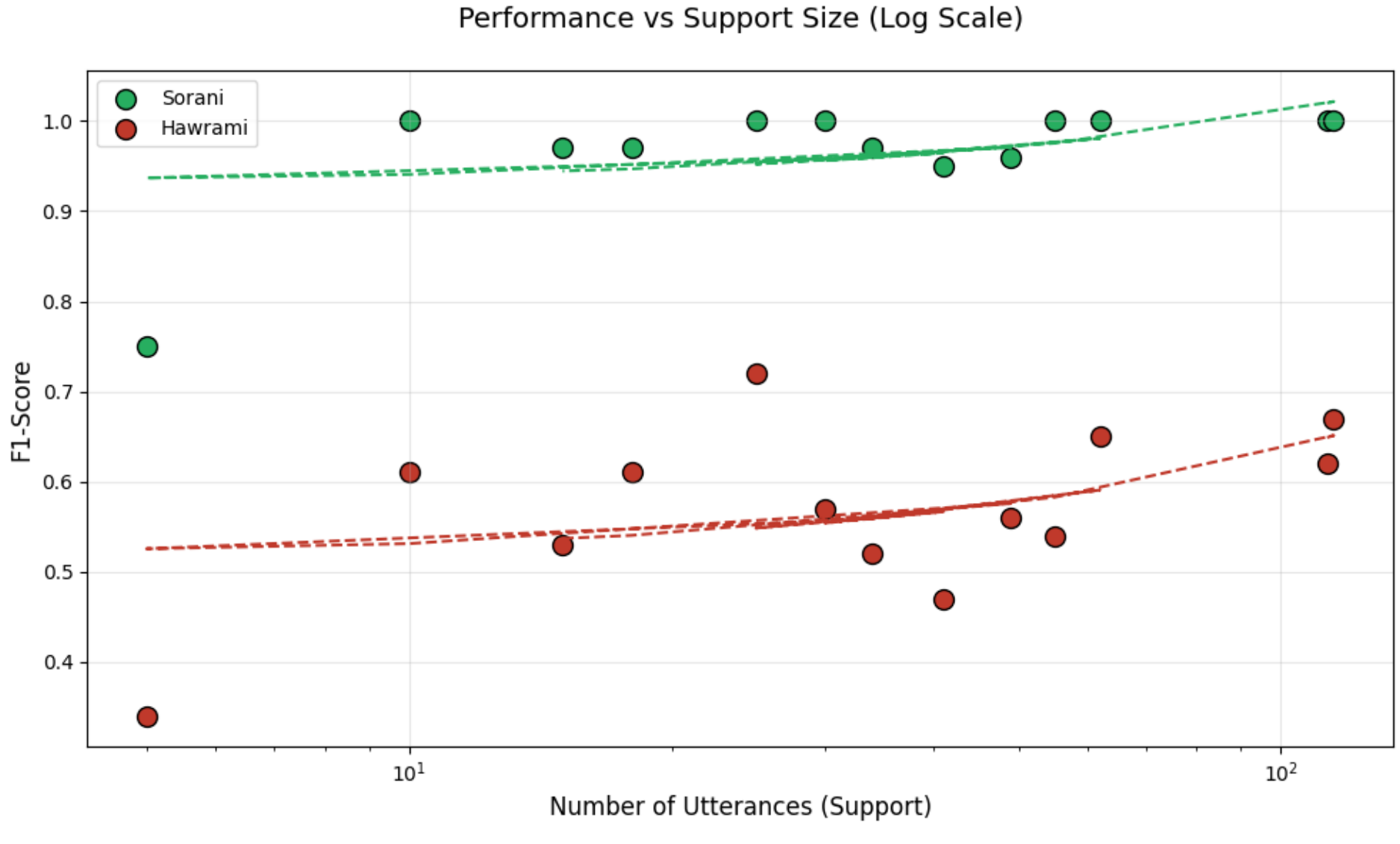

*Figure 13.Performance vs. Support Size*

Figure 13 demonstrates how recognition performance relates to the number of utterances. The Sorani dialect (green) shows remarkably consistent high performance across varying support sizes. In contrast, the Hawrami dialect (Bimbot et al.) exhibits more significant variability, with performance gradually improving as the number of utterances increases.

## Conclusion

The important linguistic variety and technological difficulties presented by many Kurdish dialects—Sorani, Kurmanji, and Hawrami—are highlighted in this research on Kurdish speaker recognition. The results highlight significant differences in identification accuracy across different speakers, therefore stressing the requirement of sophisticated approaches to handle this complexity. With a remarkable recognition rate of 0.97, Sorani is the standard dialect; Kurmanji and Hawrami behind with scores of 0.77 and 0.58, respectively. This performance difference captures the subtle differences of Kurdish language. It is abundantly evident from the great obstacles to cross-dialect adaptation that more complex technologies are required. Furthermore, performance mappings show that whereas Sorani keeps constant accuracy, Hawrami's performance increases with more utterances, implying possible improvement via careful data collecting and machine learning. Our studies show that present technologies have restrictions, so adaptive strategies become even more important. Advanced methods including transfer learning show promise for addressing these obstacles. This study generally clarifies the difficulties of Kurdish speaker detection as well as emphasizes the need of multidisciplinary approaches combining technology with cultural knowledge. We open the path for more complex, culturally sensitive voice recognition methods that fit various language environments by considering dialectal variation.